\documentclass[aps,twocolumn,superscriptaddress]{revtex4}
\usepackage{amsmath}
\usepackage{amssymb}
\usepackage{amsthm}
\usepackage{amsfonts}
\usepackage{listings}
\usepackage{enumerate}
\usepackage{latexsym}
\usepackage{color}
\usepackage{bm}
\usepackage{hyperref}
\hypersetup{
 pdfnewwindow=true, colorlinks=true,
 linkcolor=blue, anchorcolor=blue,
 citecolor=blue, filecolor=blue,
 menucolor=blue, urlcolor=blue}

\usepackage{psfrag}

\usepackage{float}
\usepackage{bm}
\usepackage{graphicx}
\usepackage{subfigure}
\usepackage{epstopdf}
\usepackage{tikz-cd}

\RequirePackage[normalem]{ulem} 
\RequirePackage{color}\definecolor{RED}{rgb}{1,0,0}\definecolor{BLUE}{rgb}{0,0,1} 

\newcommand{\beq}{\begin{equation}}
\newcommand{\eneq}{\end{equation}}

\newcommand{\bra}[1]{\left\langle#1\right|}
\newcommand{\ket}[1]{\left|#1\right\rangle}

\newcommand{\wt}[1]{\widetilde #1}

\newcommand{\bk}{{\bf k}}

\begin{document}

\def\CuBr{Cu$_2$Te$_2$O$_5$Br$_2$}
\def\HS{Hg$_2$SnSe$_4$}
\def\PS{InPS$_4$}
\def\mag{Cu$_2$MnSnSe$_4$}
\def\ie{{\it i.e.},\ }
\def\eg{{\it e.g.}\ }
\def\ea{{\it et al.}}
\input{epsf}

\tolerance 10000

\draft

\title{High-throughput screening for Weyl Semimetals with S$_4$ Symmetry}

\author{Jiacheng Gao}
\affiliation{Beijing National Laboratory for Condensed Matter Physics,
and Institute of Physics, Chinese Academy of Sciences, Beijing 100190, China}
\affiliation{University of Chinese Academy of Sciences, Beijing 100049, China}

\author{Yuting Qian}
\affiliation{Beijing National Laboratory for Condensed Matter Physics,
and Institute of Physics, Chinese Academy of Sciences, Beijing 100190, China}
\affiliation{University of Chinese Academy of Sciences, Beijing 100049, China}

\author{Simin Nie}
\email{smnie@stanford.edu}
\affiliation{Department of Materials Science and Engineering, Stanford University, Stanford, CA 94305, USA}

\author{Zhong Fang}
\affiliation{Beijing National Laboratory for Condensed Matter Physics,
and Institute of Physics, Chinese Academy of Sciences, Beijing 100190, China}
\affiliation{University of Chinese Academy of Sciences, Beijing 100049, China}

\author{Hongming Weng}
\affiliation{Beijing National Laboratory for Condensed Matter Physics,
and Institute of Physics, Chinese Academy of Sciences, Beijing 100190, China}
\affiliation{University of Chinese Academy of Sciences, Beijing 100049, China}

\author{Zhijun Wang}
\email{wzj@iphy.ac.cn}
\affiliation{Beijing National Laboratory for Condensed Matter Physics,
and Institute of Physics, Chinese Academy of Sciences, Beijing 100190, China}
\affiliation{University of Chinese Academy of Sciences, Beijing 100049, China}

\begin{abstract}
Based on irreducible representations (or symmetry eigenvalues) and compatibility relations (CR), a material can be predicted to be a topological/trivial insulator (satisfying CR) or a topological semimetal (violating CR). However, Weyl semimetals (WSMs) usually go  beyond this symmetry-based strategy. In other words, Weyl nodes could emerge in a material, no matter if its occupied bands satisfy CR, or if the symmetry indicators are zero.
In this work, we propose a new topological invariant $\chi$ for the systems with S$_4$ symmetry
(i.e., the improper rotation S$_4~(\equiv IC_{4z})$ is a proper four-fold rotation ($C_{4z}$) followed by inversion ($I$)), which can be used to diagnose the WSM phase. Moreover, $\chi$ can be easily computed through the one-dimensional  Wilson-loop technique.
By applying this method to the high-throughput screening in our first-principles calculations, we predict a lot of WSMs in both nonmagnetic and magnetic compounds. Various interesting properties (e.g., magnetic frustration effects, superconductivity and spin-glass order, etc.) are found in predicted WSMs, which provide realistic platforms for future experimental study of the interplay between Weyl fermions and other exotic states.
\end{abstract}

\maketitle
\section{Introduction}
In the last few years, the understanding of topological phases has been greatly improved due to the significant progresses in theoretical algorithms, such as symmetry classifications~\cite{po2017symmetry,khalaf2018symmetry,SlagerPRX2017}, symmetry indicators (SIs)~\cite{song2018quantitative,tang2019efficient,ono2018unified}, and topological quantum chemistry (TQC)~\cite{bradlyn2017topological,cano2018building,vergniory2017graph,cano2018topology}. These theories are very similar in identifying topological phases in three-dimensional (3D) crystals. Additionally, the fragile phases~\cite{po2018fragile} can be diagnosed by the TQC. In principle, they all rely on the same symmetry-based strategy, in which two ingredients are needed. The first one is the irreducible representations (irreps) of occupied states in a material, which can be computed
with the open-source code $irvsp$~\cite{gao2020irvsp,aroyo2011crystallography,stokes2013tabulation}. The other one is the CR in 230 space groups (SGs), which are released for the first time in the work of TQC~\cite{bradlyn2017topological} and open-accessible on the Bilbao Crystallographic Server (BCS)~\cite{aroyo2011crystallography,stokes2013tabulation}. With the code -- $CheckTopologicalMat$ -- on the BCS~\cite{vergniory2019complete}, the topology of a material can be ``automatically'' diagnosed based on the occupied states at several maximal high-symmetry $k$-points in the 3D Brillouin zone (BZ).
Recently, the high-throughput screening of topological materials has been performed for nonmagnetic compounds~\cite{tang2019comprehensive,zhang2019catalogue,vergniory2019complete}.

Generally speaking, if the occupied bands of a material satisfy CR, it would be classified as an insulator. Otherwise, it is classified as a symmetry-enforced semimetal. However, it is well known that Weyl nodes~\cite{murakami2007phase,wan2011topological,burkov2011weyl,liu2014weyl,weng2015weyl,soluyanov2015type,nie2017topological,nie2019magnetic} do not need any symmetry protections (but the lattice translations), making the appearance of Weyl nodes usually beyond the symmetry-based strategy. In other words, no matter if its occupied bands satisfy the CR, or if the SIs are zero, Weyl nodes can still emerge in the system. Two conjectures are given as below: (1) the ``trivial insulators'' classified by the symmetry-based strategy (i.e., all SIs are zero) could be WSMs; (2) the ``topological insulators'' predicted by the nonzero SIs could also be WSMs~\cite{tang2019comprehensive,zhang2019catalogue,vergniory2019complete}.

Similar to the SIs for topological (crystalline) insulators~\cite{po2017symmetry,song2018quantitative}, we here present an effective method to diagnose the WSM phase in the systems with S$_4$ symmetry~\cite{cai2015single}, no matter whether time-reversal symmetry (TRS) is respected or not.
A new topological invariant $\chi$ is defined as $1/\pi$ times the integral of the Berry curvature on a certain surface in the 3D BZ. By employing the one-dimensional (1D) Wilson-loop technique, $\chi$ can be easily computed by tracing the evolution of the average Wannier charge centers (WCC).
We have demonstrated that a  nonzero  $\chi$ guarantees the existence of Weyl nodes in an S$_4$-invariant system. By applying the new method to our first-principles calculations on the materials of 20 SGs, whose symmetry operators contain S$_4$ symmetry but no inversion symmetry ($I$), we predict a lot of WSMs in both nonmagnetic and magnetic compounds, in which many interesting properties (e.g., magnetic frustration effects, superconductivity or spin glass order, etc.) are observed. The predicted WSMs provide realistic platforms for future experimental study of the interplay between Weyl fermions and other exotic states.

\section{The definition and physical meaning of the topological invariant $\chi$ }
\subsection{Systems with S$_4$ symmetry}
\begin{figure}[!tb]
\centering
\includegraphics[width=7.5 cm,angle=0]{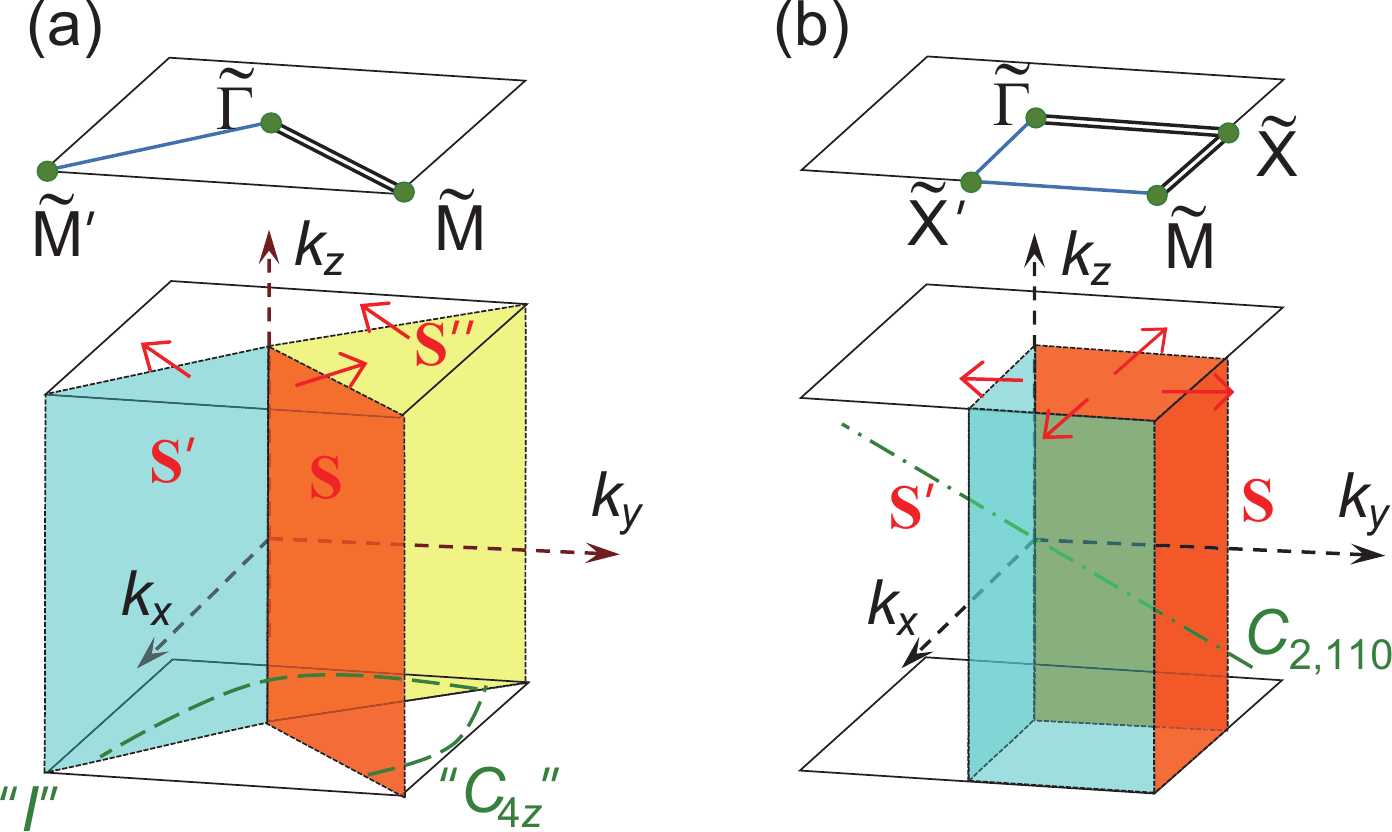}
\caption{(Color online)
The topological invariant $\chi$ is defined as $1/\pi$ times the integral of the Berry curvature on $\bf S$ surface, which is colored in orange.
$\bf S$ surface is spanned by the $z$ axis and a $xy$-plane path, whose normal vector is depicted by the red arrow. The height of $\bf S$ surface is a reciprocal lattice vector in the $k_z$ axis, which is along the rotational axis of the improper rotation  S$_{4}$. The (001) surface BZ  is shown on the top of the bulk BZ. The $xy$-plane paths are indicated by double lines in the (001) surface BZ: $\wt \Gamma-{\rm \wt M}$ in (a) and $\wt \Gamma-{\rm \wt X} -{\rm \wt M}$ in (b).
} \label{fig:WLpath}
\end{figure}

In a general case, we consider only S$_4$ symmetry in a system.
In order to determine the existence of Weyl nodes in an S$_4$-symmetric system, one could compute the net topological charge (NTC) enclosed in a quarter BZ, reshaped as a triangular prism (spanned by the $z~(k_z)$ axis and a $xy$-plane ($k_xk_y$-plane) triangle ${\rm \wt M'}$--$\wt \Gamma-{\rm \wt M}-{\rm \wt M'}$), which can be expressed as
\beq
2\pi\times {\rm NTC} =\int_{\bf S'} {\bf \Omega} {\,\rm d}{\bf S'}
   +\int_{\bf S} {\bf \Omega} {\,\rm d}{\bf S}
   +\int_{\bf S_3} {\bf \Omega} {\,\rm d}{\bf S_3},
\label{eq:NTC}
\eneq
where the vector $\bf \Omega$ is the Berry curvature with three components in 3D momentum space.
It is worth noting that $\bf S$ ($\bf{S'}$) surface must be fully gapped to keep the integral well defined on the surface. In Fig.~\ref{fig:WLpath}a, the two surfaces ($\bf S$ and $\bf{S'}$) of the prism are colored in orange and cyan, respectively, while $\bf S_3$ surface spanned by the $z$ axis and ${\rm \wt M}-{\rm \wt M'}$ segment is not shadowed. Their normal vectors are pointing from inside to outside of the prism.
The third integral on $\bf S_3$ surface has to be zero in Eq.~(\ref{eq:NTC}), since $C_{2z}$ symmetry (i.e., $C_{2z}\equiv[S_4]^2$) reverses the normal vector of this surface.

To reveal the relationship between the first and the second terms under S$_4$ symmetry, we introduce an intermediate surface $\bf S''$ (colored in yellow in Fig.~\ref{fig:WLpath}a).
The $C_{4z}$ symmetry and $I$ yield the following transformations
\[
  \begin{tikzcd}[column sep=15pt]
    \int_{\bf S} {\bf \Omega} {\,\rm d}{\bf S} \ar[rr,equal,"S_{4}",shorten < =2mm,shorten > =2mm]  \ar[dr,"C_{4z}" swap] & & \int_{\bf S'} {\bf \Omega} {\,\rm d}{\bf S'}  \\
   & \int_{\bf S''} {\bf \Omega} {\,\rm d}{\bf S''} \ar[ur,"I" swap]&
  \end{tikzcd}
 \]
where the defined normal vector of $\bf S''$ surface is depicted in Fig.~\ref{fig:WLpath}a.
The first transformation is because of $C_{4z}$ symmetry, and the second transformation is because $I$ makes $\bf \Omega(k)=\Omega(-k)$. Therefore, under the combined S$_{4}$ symmetry , the first integral is equal to the second one in the right-hand part of Eq.~(\ref{eq:NTC}).
Note that neither $C_{4z}$ nor $I$ is preserved in these systems.

Then, we define a topological invariant
\beq
\chi\equiv \frac{1}{\pi} \int_{\bf S} {\bf \Omega} {\,\rm d}{\bf S}.
\label{eq:chi}
\eneq
Substituting Eq.~(\ref{eq:chi}) into Eq.~(\ref{eq:NTC}), the NTC can be simply expressed as ${\rm NTC}=\chi$.
It is worth noting that the topological invariant $\chi$ is an integer, which indicates the NTC in a quarter of BZ (i.e., the prism in Fig.~\ref{fig:WLpath}a). Later, we will show that $\chi$ can be easily calculated with the 1D Wilson-loop technique.

\begin{figure}[!tb]
\centering
\includegraphics[width=8.5 cm,angle=0]{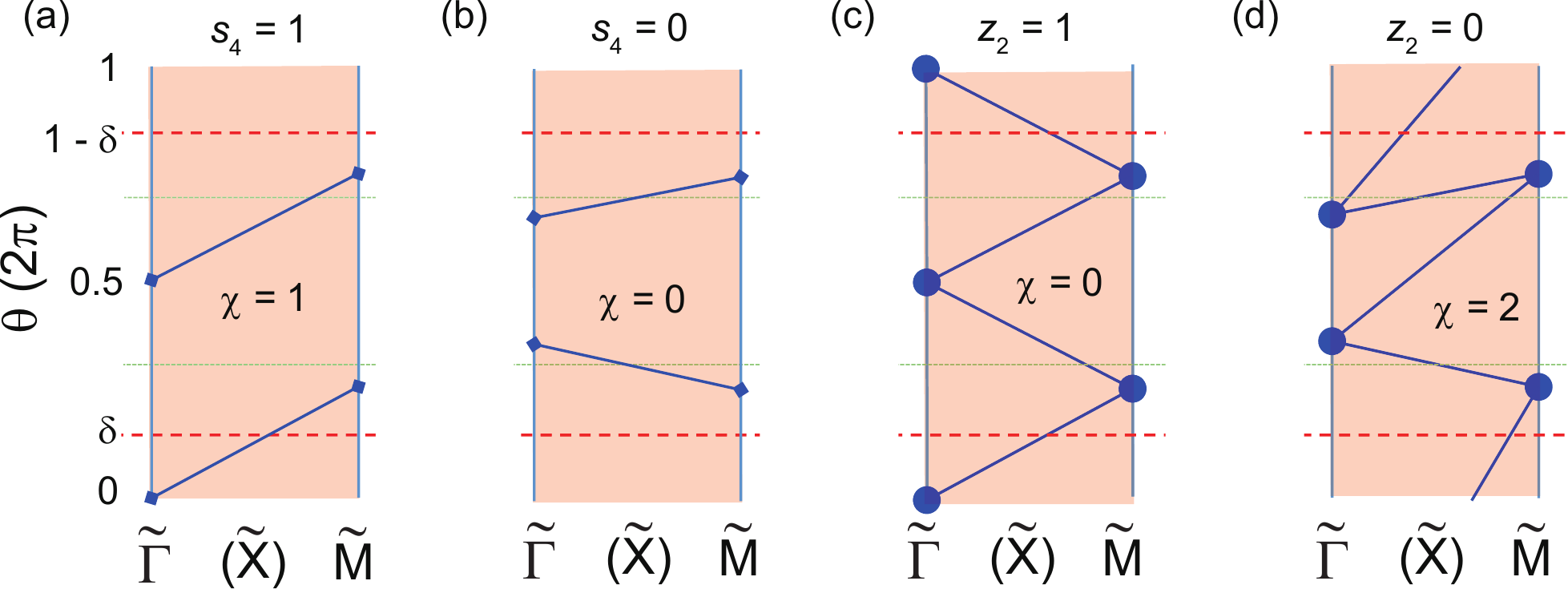}
\caption{(Color online)
Defined as the integral of the Berry curvature on $\bf S$ surface in 
Fig.~\ref{fig:WLpath}, the topological invariant $\chi$ can be easily 
obtained in the Wilson-band plots along the $xy$-plane path 
($\wt \Gamma-{\rm \wt M}$ or $\wt \Gamma -{\rm \wt X} -{\rm \wt M}$),
 by counting the number of the positively-sloped bands crossing two 
 horizontal reference (red-colored dashed) lines 
 ($\theta=2\pi\delta $ and $\theta=2\pi (1-\delta)$ with $\delta\neq 0$ or 0.5), and subtracting from it the number of negatively-sloped crossing ones.
The individual phases ($\theta$) of Wilson eigenvalues are plotted for a model with two (four) occupied bands for magnetic (nonmagnetic) system.
Panels (a) and (b) don't have TRS, while panels (c) and (d) have TRS. Filled circles stand for double degeneracies.
$s_4$ is defined in Eq. (5), and $z_2$ is defined in Ref.~\cite{song2018quantitative} (and Eq. (18) in the Supplementary materials).
In the absence of TRS, $\chi=2n+1$ with $s_4=1$ and $\chi=2n$ with $s_4=0$.
} \label{fig:WLband}
\end{figure}

\subsection{S$_4$-invariant systems with $C_{2,110}$ symmetry}
\label{s4def}
In 230 SGs, we find that there are 20 SGs, which have S$_4$ symmetry, but neither four-fold rotational symmetry nor $I$. These S$_4$-invariant SGs are listed in Table~\ref{tab:mat}.
The invariant $\chi$ defined above is applicable to most of them, except for eight ones.
These exceptional SGs have a two-fold rotational symmetry along the [110] direction (i.e., $C_{2,110}$) as shown in Fig.~\ref{fig:WLpath}b, which makes the previous $\chi$ always being zero. However, after re-defining the $\bf S$ surface with the $xy$-plane path $\wt \Gamma-{\rm \wt X}-{\rm \wt M}$ (shown as double lines in Fig.~\ref{fig:WLpath}b), a new invariant $\chi$ can be defined. With S$_4$ symmetry, one can have
\beq
\int_{\bf S} {\bf \Omega} {\,\rm d}{\bf S}=\int_{\vec S} {\bf \Omega} {\,\rm d}{\vec S}~(+\int_{\bf S_3} {\bf \Omega} {\,\rm d}{\bf S_3})=\int_{\bf S'} {\bf \Omega} {\,\rm d}{\bf S'},
\eneq
where the surface $\vec S$ is spanned by the $k_z$ axis and $xy$-plane path $\wt \Gamma-{\rm \wt X'}-{\rm \wt M'}$ in Fig.~\ref{fig:WLpath}b.
With the re-defined  $\bf S$ surface in Fig.~\ref{fig:WLpath}b, the new invariant $\chi$ in Eq.~(\ref{eq:chi}) still indicates the NTC in the re-defined quarter of BZ (i.e., the tetragonal prism in Fig.~\ref{fig:WLpath}b).
More generally, the invariant $\chi$ can be defined on a $k_z$-directed surface with  any  $xy$-plane path starting from $\wt \Gamma$ and ending at ${\rm \wt M}$.
It is also noted that our strategy to find WSMs in S$_4$-invariant systems can be generalized to the systems with other crystalline symmetries, which are not the main topic here and left in the future work.

\begin{table*}[!htb]
  \caption{
  A list of WSMs with the  nonzero  topological invariant $\chi$ in our first-principles calculations.
  The calculations for most of them are performed in the nonmagnetic state. The asterisks indicate the calculations in the magnetic state.
  NA implies that there are crossing points between conduction bands and valence bands along the $k_z$-directed line through $\Gamma$ or M in the WSM.
  The Wilson bands and electronic band structures of these materials are given in the Supplementary materials.
  }\label{tab:mat}
  \begin{tabular}{ccccclr  p{1.0cm}  ccccclr}
  \hline
  SG & Symm. & BZ & WLP  &$\chi$\ \ & Formula & \#ICSD         && SG & Symm. & BZ & WLP                        &$\chi$& Formula & \#ICSD\\
  &     & Fig.S2 & Fig. \ref{fig:WLpath}    &    &         &                &&    &     & Fig.S2 & Fig. \ref{fig:WLpath}    &    &         &       \\
  \cline{1-7}
  \cline{9-15}
  81 & P-$4$ & (a) & (a) & $-$2\ \ & Cu$_2$Te$_2$O$_5$Br$_2$       & \#152959       && 121 & I-$42m$ & (b) & (a) &  2\ \  & $^\ast$Fe$_2$AgGaTe$_4$    & \#192470 \\
     &       &     &     & $-$2\ \  & Cu$_2$Te$_2$O$_5$Cl$_2$       & \#89978        &&     &         &     &     & $-$1\ \  & $^\ast$Cu$_2$MnSnSe$_4$    & \#155904 \\
  \cline{1-7}
  82 & I-$4$ & (b) & (a) &  2\ \  & InPS$_4$                      & \#23612        &&     &         &     &     & $-$2\ \  & $^\ast$Cu$_2$MnSnS$_4$     & \#193982 \\
     &       &     &     & $-$2\ \  & Hg$_2$SnSe$_4$~$~$            & \#639205       &&     &         &     &     & NA\ \ & $^\ast$Ag$_2$FeSnS$_4$     & \#42534 \\
     &       &     &     & NA\ \ & $^\ast$Fe$_2$NiP              & \#153485       &&     &         &     &     & NA\ \ & $^\ast$Cu$_2$FeSnSe$_4$    & \#85126  \\
     &       &     &     & NA\ \ & $^\ast$Fe$_2$NiB              & \#614131       &&     &         &     &     & NA\ \ & $^\ast$Cu$_2$CoSnSe$_4$    & \#99296 \\
     &       &     &     &  2\ \  & $^\ast$Ni$_3$P                & \#27161        &&     &         &     &     & NA\ \ & $^\ast$GaMn$_2$Se$_4$      & \#190460  \\
     &       &     &     & $-$4\ \  & FeGa$_2$S$_4$                 & \#602024       &&     &         &     &     &  2\ \  & Cu$_2$CdGeTe$_4$           & \#165094  \\
     &       &     &     & $-$2\ \  & TaK$_3$F$_8$O$_6$             & \#423165       &&     &         &     &     & $-$2\ \  & Fe$_2$CuTlAs$_2$S$_6$      & \#252986 \\
     &       &     &     & $-$4\ \  & AsBCa$_2$O$_8$                & \#27527        &&     &         &     &     & $-$2\ \  & Cu$_3$AsSe$_4$             & \#610361  \\
     &       &     &     & 10\ \  & Cr$_3$P                       & \#23560        &&     &         &     &     &  2\ \  & Cu$_2$ZnGeSe$_4$           & \#627831  \\
     &       &     &     &$-$10\ \  & Ta$_3$Ge                      & \#56027        &&     &         &     &     &  2\ \  & Cu$_2$ZnSnSe$_4$           & \#629099 \\
     &       &     &     & $-$6\ \  & Zr$_3$Sb                      & \#195057       &&     &         &     &     &  2\ \  & Cu$_2$CdSnSe$_4$           & \#619784 \\
  \cline{1-7}
  111 & P-$42m$ & (a) & (a) &  4\ \  & FeGa$_2$Se$_4$             & \#631817       &&     &         &     &     & 2\ \  & Cu$_2$HgSnSe$_4$           & \#627836 \\
  \cline{1-7}
  112 & P-$42c$ & (a) & (a) &  2\ \  & CuIn$_3$Se$_4$             & \#91493        &&     &         &     &     &  2\ \  & Cu$_2$HgGeTe$_4$~\cite{qian2019weyl} & \#627904 \\
  \cline{1-7}
                                                                               \cline{9-15}
  113 & P-$42_1m$&(a) & (a) & $-$2\ \  & MnSr$_2$Ge$_2$O$_7$        & \#84033        && 122 & I-$42d$ & (b) & (a) & 2\ \   & KPO$_4$                    & \#33583 \\
      &          &    &     &  2\ \  & CoSi$_2$Ca$_2$O$_7$        & \#186949       &&     &         &     &     &$-$4\ \   & RbPO$_4$H$_4$              & \#54871 \\
      &          &    &     & $-$2\ \  & GaCa$_2$B$_2$O$_7$         & \#162769       &&     &         &     &     &$-$2\ \   & CsAsO$_4$                  & \#44799 \\
  \cline{1-7}
  114 & P-$42_1c$&(a) & (a) & $-$2\ \  & Ag$_2$SO$_4$N$_4$          & \#36230        &&     &         &     &     & 2\ \   & Rb$_2$O$_3$                & \#248539 \\
                                                                               \cline{9-15}
      &          &    &     & $-$4\ \  & BiO$_2$                    & \#291519       && 215 & P-$43m$ & (a) & (a) &$-$2\ \   & Cu$_3$AsS$_4$              & \#42516 \\
  \cline{1-7}
                                                                               \cline{9-15}
  115 & P-$4m2$ & (a) & (b) & $-$2\ \  & Pb$_2$OF$_2$               & \#76964        && 216 & F-$43m$ & (c) & (b) & 4\ \   & GaRhLi$_2$                 & \#106714 \\
      &         &     &     &  4\ \  & Hf$_2$GaSb$_3$             & \#189076       &&     &         &     &     & 2\ \   & MgPtSn                     & \#16479 \\
  \cline{1-7}
  116 & P-$4c2$ & (a) & (b) &    & $\varnothing$              &                &&     &         &     &     &$-$2\ \   & Os$_2$CdN$_4$O$_8$         & \#25306 \\
  \cline{1-7}
  117 & P-$4b2_1$& (a) & (b) & 8\ \  & Bi$_2$O$_3$                & \#27151        &&     &         &     &     &$-$4\ \   & GeV$_4$Se$_8$              & \#195256 \\
  \cline{1-7}
                                                                               \cline{9-15}
  118 & P-$4n2_1$& (a) & (b) &$-$2\ \  & ZnSb$_2$                   & \#36165        && 217 & I-$43m$ & (d) & (a) &$-$2\ \   & Cu$_3$AsS$_3$              & \#33588 \\
  \cline{1-7}
  119 & I-$4m2$ & (b) & (b) & $-$2\ \  & AlPO$_4$                   & \#162670       &&     &         &     &     &$-$2\ \   & Co$_4$O$_{13}$B$_6$        & \#96561 \\
      &         &     &     & $-$2\ \  & AgTlTe$_2$                 & \#43284        &&     &         &     &     & 2\ \   & Na$_2$Al$_3$S$_3$O$_6$     & \#63022 \\
  \cline{1-7}
                                                                               \cline{9-15}
  120 & I-$4c2$ & (b) & (b) &    & $\varnothing$              &                && 218 & P-$43n$ & (a) & (a) &    & $\varnothing$              & \\
                                                                               \cline{9-15}
      &         &     &     &    &                            &                && 219 & F-$43c$ & (c) & (b) &    & $\varnothing$              & \\
                                                                               \cline{9-15}
      &         &     &     &    &                            &                && 220 & I-$43d$ & (d) & (a) &$-$4\ \   & La$_2$C$_3$                & \#26588 \\
      &         &     &     &    &                            &                &&     &         &     &     & 2\ \   & Y$_2$C$_3$                 & \#77572 \\
  \hline
  \end{tabular}
\end{table*}

\subsection{1D Wilson-loop technique}
To compute the integral for the topological invariant $\chi$, one can employ the 1D Wilson-loop technique.
The average WCC [$\Theta(k_x,k_y)$] of the $k_z$-directed Wilson loops [$(k_x,k_y,0)-(k_x,k_y,2\pi)$] are defined as follows
\begin{eqnarray*}
\Theta(k_x,k_y)&=&\int_{(k_x,k_y,0)}^{(k_x,k_y,2\pi)} {\rm Tr}[\vec A(\bk)] {\,\rm d}l,\\
 e^{i\Theta(k_x,k_y)}&=&{\rm Det}[W(k_x,k_y)],\\
 W_{mn}(k_x,k_y)&\equiv& \bra{u_m(k_x,k_y,2\pi)}\prod_{k_z}^{2\pi \gets 0}P(\bk) \ket{u_{n}(k_x,k_y,0)}
\end{eqnarray*}
where $\vec A(\bk)=i\bra{u_m(\bk)}\partial_\bk\ket{u_m(\bk)}$, and $\ket{u_m(\bk)}$ are the Bloch eigenstates of the Hamiltonian $H(\bk)$ of the system. The indices $m, n$ run over its occupied bands,  and $P(\bk) = \ket{u_m(\bk)}\bra{u_m(\bk)}$ is the projector on the subspace of occupied bands at momentum $\bk$. 
It is also known as $\Theta(k_x,k_y)=\sum_j\theta_j(k_x,k_y)$, with $\theta_j(k_x,k_y)$ representing the phase of the $j$th eigenvalue of the Wilson matrix $W(k_x,k_y)$ on occupied Bloch bands.

By plotting the average WCC $\Theta(k_x,k_y)$ (or individual phases $\theta_j(k_x,k_y)$) as a function of the $xy$-plane path (called ``Wilson bands" for short), the Eq.~(\ref{eq:chi}) can be rewritten as
\beq
\chi=\frac{1}{\pi}\int_{\wt\Gamma-{\rm \wt M}~or~\wt \Gamma- {\rm \wt X}-{\rm \wt M}}{\,\rm d}\Theta(k_x,k_y).
\label{eq:theta}
\eneq
The $xy$-plane paths for S$_4$-invariant SGs in our calculations are given in Table~\ref{tab:mat}.
Thus, regardless of which path of the Wilson loops (WLP) in Fig.~\ref{fig:WLpath} used in our calculations, the $\chi$ can be easily obtained in a Wilson-band plot, by counting the number of the positively-sloped bands crossing two horizontal reference lines ($\theta=2\pi\delta $ and $\theta=2\pi (1-\delta)$ with $\delta\neq 0$ or 1/2), and subtracting from it the number of negatively-sloped crossing ones (see examples in Fig.~\ref{fig:WLband}).

There are two particular loops at $\wt \Gamma$ and ${\rm \wt M}$ (i.e., $W(\alpha,\alpha)$ with $\alpha=0$ and $\pi$), which are the starting and ending loops, respectively.
In the presence of S$_4$ symmetry, $W(\alpha,\alpha) = W^\dagger(\alpha,\alpha)$ because S$_4$ symmetry flips the direction of the two 1D Wilson loops. This implies $\Theta(\alpha,\alpha)=-\Theta(\alpha,\alpha)~{\rm mod}~2\pi$, resulting in
$\Theta(\alpha,\alpha)=n\pi$. Substituting it into Eq.~(\ref{eq:theta}), one can conclude that $\chi$ has to be an integer. Here, we define a new S$_4$ SI in the absence of TRS below
\beq
\begin{split}
s_4&=[\Theta(0,0)+\Theta(\pi,\pi)]/\pi~{\rm mod}~2 \\
   &= \sum_{K\in {\rm SIM}} (n_K^0+n_K^1) ~{\rm mod}~2,
\end{split}
\eneq
where $n_K^\alpha$ labels the number of occupied states with S$_4$ eigenvalue $\lambda_\alpha =e^{i\pi\frac{2\alpha-1}{4}}~(\alpha=0,1,2,3)$  at the S$_4$ invariant momentum (SIM) $K$ (See more details in the  Supplementary materials).
Since $s_4=1$ implies $\chi=2n+1$ (nonzero), a magnetic WSM phase can be indicated by $s_4=1$.

\subsection{S$_4$-invariant systems with TRS}
Once imposing TRS in the systems, both S$_4$ symmetry and TRS are preserved on the two particular $k_z$-directed loops.
More importantly, each $k$-point on the loops respects $C_{2z}$ symmetry, which makes the two Wilson matrices ($W(\alpha,\alpha), \alpha=0,\pi$)  block-diagonal by the $C_{2z}$ eigenvalues $\pm i$.
TRS sends the states in the $+i$ subspace to those in the $-i$ subspace, while S$_4$ symmetry relates them in the same subspace.
Considering $2N$ occupied bands, TRS makes the eigenvalues of the two Wilson matrices in the form of $\{e^{i\theta_2},e^{i\theta_2},e^{i\theta_4},e^{i\theta_4}, \dots, e^{i\theta_{2N}}, e^{i\theta_{2N}}\}$.
On the other hand, in one subspace (e.g., $C_{2z}$ eigenvalue $+i$), S$_4$ symmetry yields the constraint: $\phi(\alpha)=-\phi(\alpha)~{\rm mod}~2\pi$ with $e^{i\phi(\alpha)}=\prod_{j=1}^N e^{i\theta_{2j}(\alpha,\alpha)}$, resulting in $\phi(\alpha)=n\pi$.
Considering the other subspace related by TRS, one concludes that $\Theta(\alpha,\alpha)=2\phi(\alpha)=2n\pi$, leading that the invariant $\chi$ has to be an even number ($\chi=2n$) for the systems with both S$_4$ symmetry and TRS.
It is notable that the S$_4$ $z_2$ can be actually defined as $z_2=[\phi(0)+\phi(\pi)]/\pi$ mod 2 in the Wilson-loop calculations (see details in the  Supplementary materials).

\section{High-throughput screening of WSMs}
Based on the above derivations, we conclude that a nonzero $\chi$ indicates the existence of Weyl nodes in the S$_4$-invariant systems.
By computing the topological invariant $\chi$, one can easily diagnose the WSM phase in a material. Here, we sweep through most of  materials with S$_4$ symmetry in the Inorganic Crystal Structure Database (ICSD) to search for WSMs (see filter details in the Supplementary materials). In the 20 SGs with S$_4$ symmetry, we find many WSMs with trivial or nontrivial SIs~\cite{song2018quantitative,po2017symmetry}, even though they satisfy the CR.
The results are listed in Table~\ref{tab:mat}.
It is worth noting that some WSM candidates may be missed with the specific $\chi=0$ used in our calculations, which could be solved by re-defining $\chi$ as mentioned in the end of Sec~\ref{s4def}.
Compared with established methods for WSM search~\cite{ivanov2019monopole,xu2020comprehensive}, our method does not need the construction of the Wannier-function based Hamiltonian model, which is time-consuming and hard to implement in high-throughput screening scheme. Moreover, instead of searching Weyl nodes in the whole BZ with a very dense $k$-mesh, our method only needs to calculate the invariant $\chi$, which is efficient and easy to implement.
\\
\indent Our first-principles calculations were performed with the VASP package \cite{KRESSE199615,vasp} based on the density functional theory (DFT) with the projector augmented wave method \cite{paw1,paw2}.
 The generalized gradient approximation with exchange-correlation functional of Perdew, Burke and Ernzerhof for  the exchange-correlation functional \cite{pbe} was employed. The cut-off energy of plane wave basis set was set to be 125\% ENMAX value in the pseudo-potential file. A $\Gamma$-centered Monkhorst-Pack grid with 30 k-points per 1/\AA~was used for the self-consistent calculations. The lattice and atomic parameters in the ICSD were employed in our calculations~\cite{monkhorst1976special}.
Spin-orbit coupling was taken into account in our first-principles calculations. The Wilson-loop technique~\cite{Yu2011An} was used to calculate topological invariants and chiral charges~\cite{Fang2003The,wan2011topological}, which was homemade and implemented within the VASP package (i.e., \emph{vasp.5.3.3}). In our magnetic calculations, the rotationally invariant DFT+U method introduced by Liechtenstein \ea~\cite{ldau} was employed to treat the electron correlation of $3d$ states. The effective on-site Coulomb interactions U$_{\text{eff}}$ on $3d$ electrons was taken to be 3 eV. The Fermi arc states were calculated based on the Green's function method~\cite{sancho1985highly} of the semi-infinite systems, which were constructed by the maximally localized Wannier functions~\cite{mlwf}.

\begin{figure}[!t]
\includegraphics[width=8.3cm]{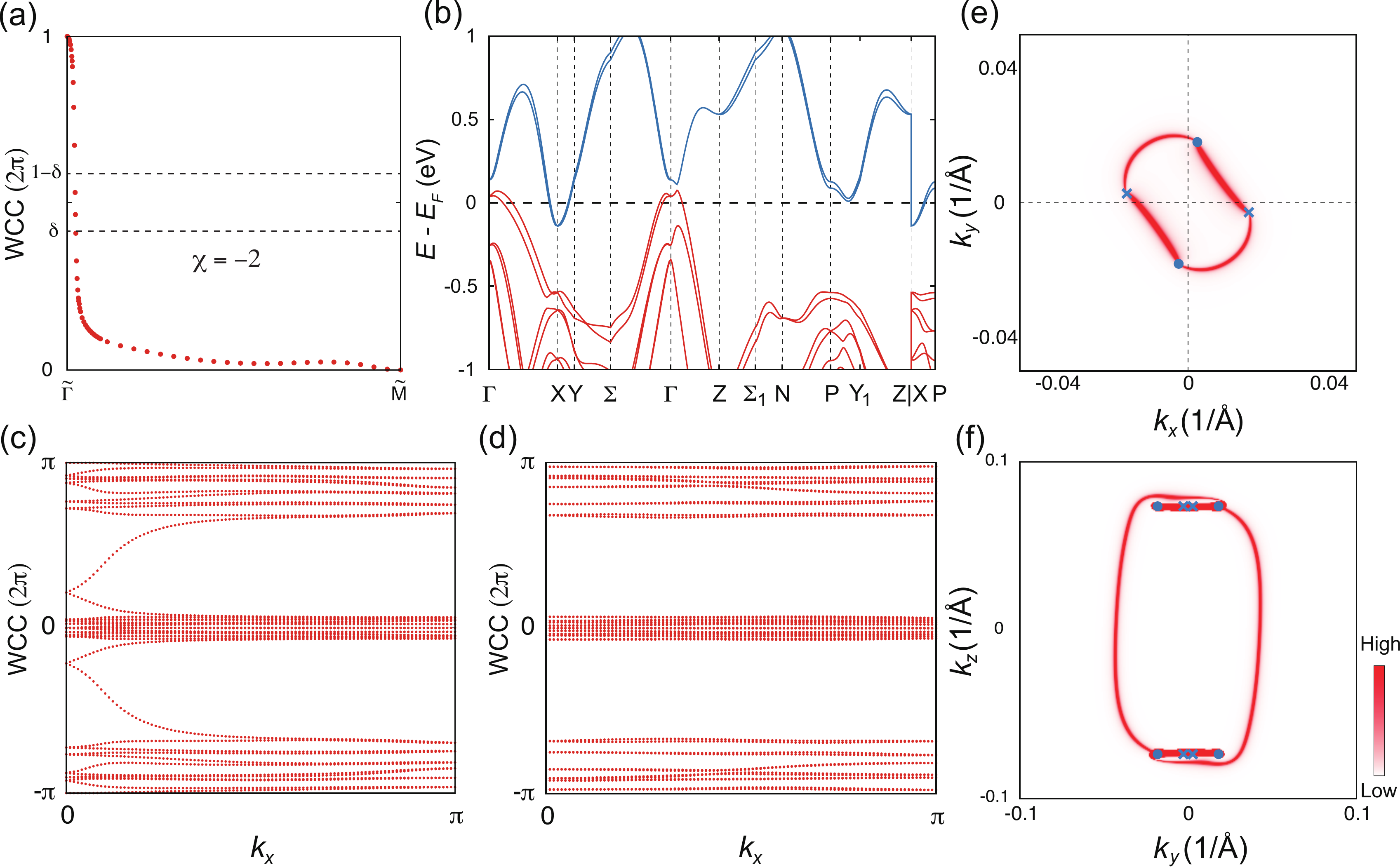}
\caption{(Color online)
(a) The Wilson bands of the $xy$-plane path of Hg$_2$SnSe$_4$. (b) The band structure of Hg$_2$SnSe$_4$.
The dashed horizontal line denotes the Fermi level ($E_{\rm F}$).
(c) and (d) The WCC as a function of $k_x$ in $k_z=0$ and $k_z=\pi$ planes, respectively. (e) and (f) The (001) and (100) surface states of Hg$_2$SnSe$_4$
at $E-E_{\rm F}=0.0945$ eV, respectively. Hereafter, the Weyl points with different chirality are shown as crossings and circles, respectively.
} \label{fig:639205_arc}
\end{figure}

\subsubsection{The class of nonmagnetic WSMs with trivial SI}
In our findings, the obtained nonzero invariant $\chi$ suggests that some materials, previously predicted to be trivial insulators according to the S$_4$ SI $z_2=0$, turn out to be WSMs. In the main text, we take Hg$_2$SnSe$_4$ and Cu$_2$Te$_2$O$_5X_2$ ($X=$Cl, Br) as two examples in this class of WSMs, which can be further checked in future experiments.

The Hg$_2$SnSe$_4$ compound crystallizes in the cadmium thiogallate structure with SG \#82~\cite{HgSnSecrystal}. Its band structure is presented in Fig.~\ref{fig:639205_arc}b, exhibiting a direct band gap along the high-symmetry lines. The occupied bands satisfy the CR and have a trivial SI (i.e., $z_2=0$). Therefore, Hg$_2$SnSe$_4$ is previously classified as a trivial insulator~\cite{zhang2019catalogue,tang2019comprehensive,vergniory2019complete}.
However, our $k_z$-directed Wilson-loop calculations along $\wt \Gamma$--$\wt M$ in Fig.~\ref{fig:639205_arc}a show that the topological invariant $\chi$ is $-2$, indicating the existence of Weyl nodes in Hg$_2$SnSe$_4$, which is consistent with the very recent work~\cite{xu2020comprehensive}.

For the systems with both S$_4$ symmetry and TRS, the WSM phase can be also checked with the criterion $\eta\neq z_2$. In Ref.~\cite{qian2019weyl}, the $\eta$ is defined as $ (-1)^{\eta} = (-1)^{\nu_{1}}(-1)^{\nu_{2}}$ with $\nu_{1}$ and $\nu_{2}$ the time-reversal $\mathbb{Z}_2$ invariants in $k_z=0$ and $k_z=\pi$ planes, respectively. Wilson bands for the two planes are plotted in Fig.~\ref{fig:639205_arc}c and d, respectively. One can see that $\nu_1=1$ and $\nu_2=0$, resulting in $\eta=1$. Since the S$_4$ $z_2$ is computed to be zero, the criterion of  $\eta\neq z_2$ is applicable to the WSM Hg$_2$SnSe$_4$.

After carefully checking the energy gaps in the full BZ, we do find eight Weyl nodes with topological charge $|C|=1$, as shown in Table~\ref{tab:charge}, which are all related by symmetries in the system.
As a hallmark of WSMs, the nontrivial surface states on the (001) and (100) surfaces are obtained and shown in Fig.~\ref{fig:639205_arc}e and f, respectively.
Hereafter, all surface states are calculated by the Green's function method based on the Wannier-based tight-binding models throughout the paper.
One can see that surface Fermi arc states are connecting the projections of the opposite-chirality Weyl points (denoted by filled squares and circles, respectively).
It is noted that two Fermi arcs have to go across the $k_z = 0$ line in Fig.~\ref{fig:639205_arc}f, since it is the edge of the nontrivial $k_z=0$ plane ($\nu_1=1$).

\begin{table}[h]
  \caption{Distribution of nonequivalent Weyl points and associated topological charges for Hg$_2$SnSe$_4$, \CuBr, Zr$_3$Sb, Ta$_3$Ge and \mag. The positions of Weyl points are given in Cartesian coordinates. The other equivalent Weyl nodes can be obtained by the crystalline symmetries.}
  \label{tab:charge}
  \begin{tabular}{ccccc}
    \hline
 WSMs& Positions ($k_x$,~$k_y$,~$k_z$) & Cha- & $E-E_{\rm F}$  & Multi- \\
     & (1/\AA) &             rge  & (eV)   & plicity \\
    \hline
Hg$_2$SnSe$_4$ &(0.0028,~0.0181,~0.0737)  &$-$1 & 0.0945 & 8\\
    \hline
    \CuBr & ( 0.3674,~0.3901,~0.4846) &$-$1 & 0.1402 & 8\\
          & ( 0.2952,~0.3460,~0.2522) &$-$1 & 0.0084 & 8\\
          & ($-$0.0947,~0.1980,~0.4937) & 1 & 0.0989 & 4\\
          & ($-$0.0497,~0.2040,~0.4937) & 1 & 0.1140 & 4\\
    \hline
    Zr$_3$Sb & ( 0.1943, 0.2129, 0.4411) &$-$1 & 0.0906 & 8\\
             & ( 0.0192, 0.3871, 0.0000) & 1 &$-$0.0338 & 4\\
             & ( 0.0493, 0.3975, 0.0000) & 1 &$-$0.0344 & 4\\
             & ($-$0.1223, 0.3743, 0.2395) &$-$1 &$-$0.0473 & 8\\
             & ($-$0.1228, 0.3640, 0.2358) &$-$1 &$-$0.0481 & 8\\
             & ($-$0.0240, 0.2715, 0.2307) &$-$1 &$-$0.0520 & 8\\
             & ($-$0.1585, 0.2538, 0.2368) & 1 &$-$0.0618 & 8\\
             & ($-$0.0706, 0.2514, 0.2362) &$-$1 &$-$0.0673 & 8\\
    \hline
    Ta$_3$Ge & ( 0.0107, 0.4512, 0.2281) &$-$1 & 0.0947 & 8\\
             & ( 0.0679, 0.2533, 0.0798) &$-$1 & 0.0395 & 8\\
             & ($-$0.0702, 0.1591, 0.5455) &$-$1 & 0.0351 & 8\\
             & ( 0.0225, 0.5648, 0.1636) &$-$1 & 0.0243 & 8\\
             & ( 0.0426, 0.0735, 0.3311) &$-$1 & 0.0139 & 8\\
             & ( 0.2291, 0.2970, 0.4729) & 1 & 0.0026 & 8\\
             & ($-$0.1229, 0.2080, 0.2910) &$-$1 &$-$0.0140 & 8\\
    \hline
    \mag     & ( 0.0001, 0.0931, $-$0.0262) & 1 &$-$0.0080 & 4\\
    (FM)     & ( 0.0001, 0.0168,  0.0007) &$-$1 & 0.0057 & 4\\
             & ( 0.0001, 0.0626, $-$0.1137) &$-$1 & 0.0170 & 4\\
    \hline
    \mag     & ($-$0.0001,0.0076,$-$0.0173) &$-$1 & 0.0100 & 4\\
    (AFM)    & ( 0.0001,0.0076, 0.0173) &$-$1 & 0.0100 & 4\\
    \hline
  \end{tabular}
\end{table}

\begin{figure}[!t]
\includegraphics[width=8.3cm]{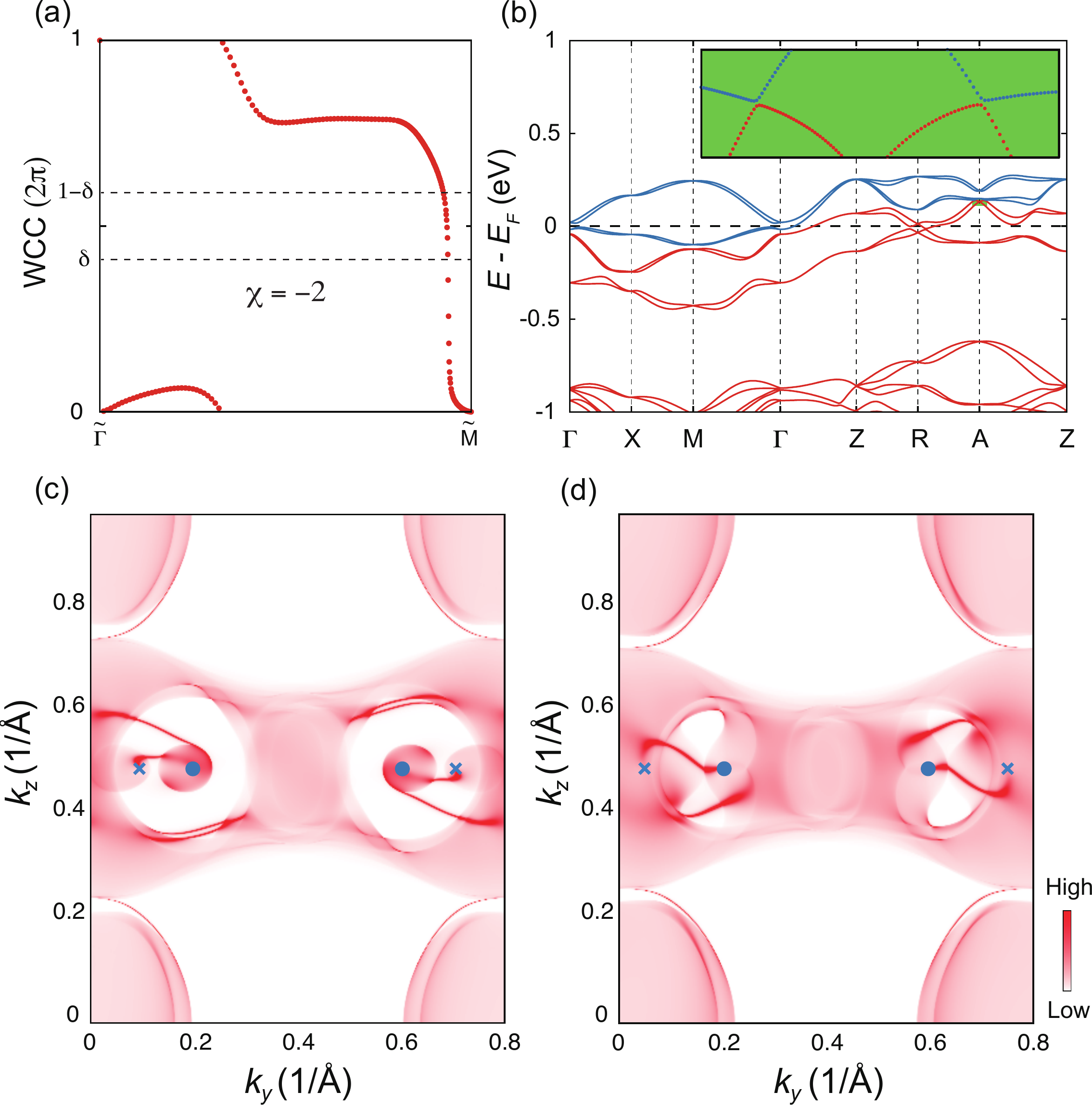}
\caption{(Color online)
(a) The Wilson bands of the $xy$-plane path of \CuBr. (b) The band structure of \CuBr.
The inset shows the zoom-in band structure around the A point. (c) and (d) The (100) surface states of \CuBr~at $E-E_{\rm F}=0.0989$ eV and $E-E_{\rm F}=0.1140$ eV, respectively. Only Weyl points of the corresponding energy are depicted.
} \label{fig:152959_arc}
\end{figure}

\begin{figure}[t]
\includegraphics[width=8.3cm]{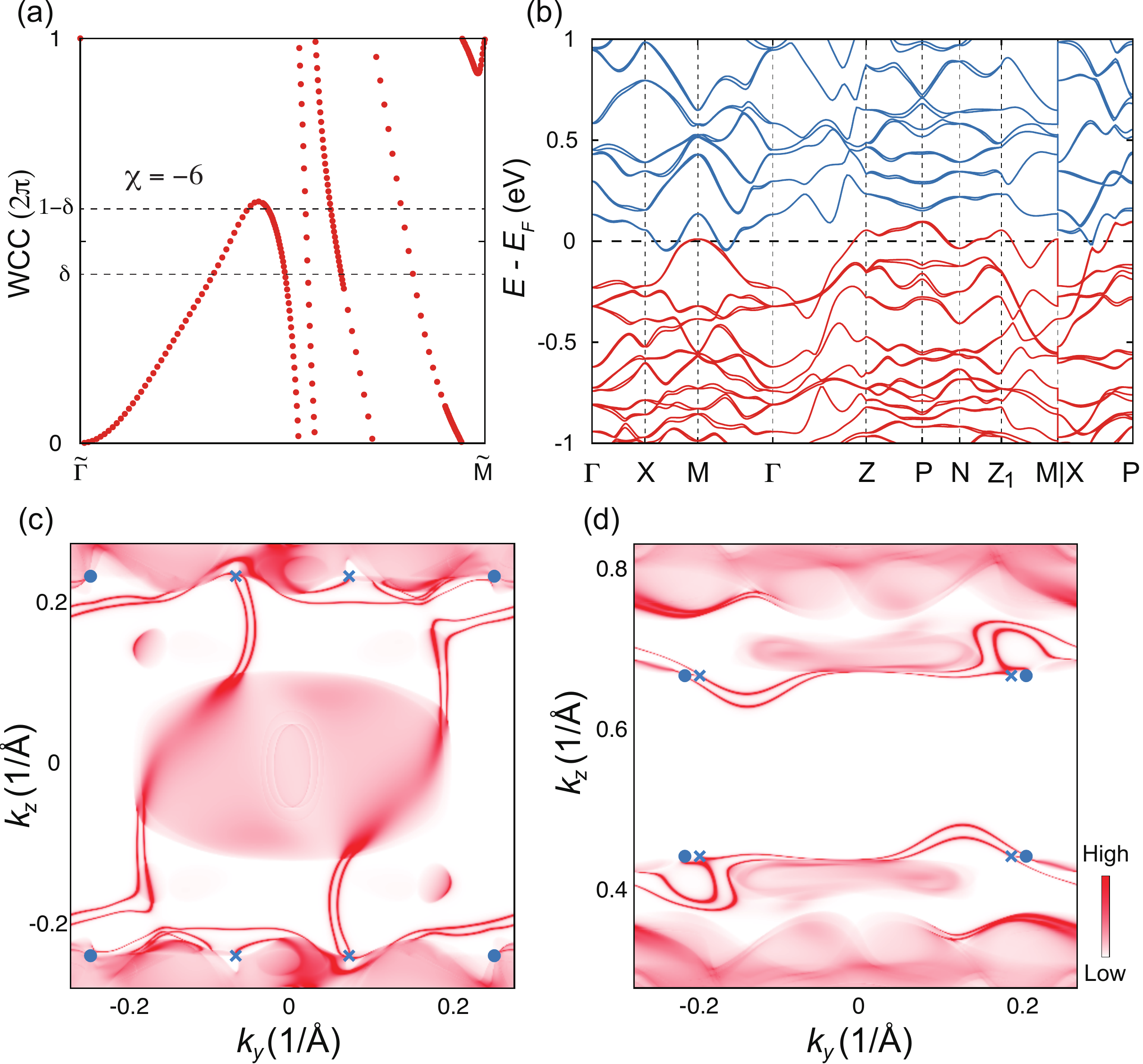}
\caption{(Color online)
(a) The Wilson bands of the $xy$-plane path of Zr$_3$Sb. (b) The band structure of Zr$_3$Sb.
(c) and (d) The (100) surface states of Zr$_3$Sb at $E-E_{\rm F}=-0.0673$ eV and $E-E_{\rm F}=0.0906$ eV, respectively. Only Weyl points of the corresponding energy are depicted.
} \label{fig:195057_arc}
\end{figure}

Next, the oxohalogenides Cu$_2$Te$_2$O$_5X_2$ ($X=$ Cl, Br) have received considerable attention in recent years, as they contain weakly coupled Cu$^{2+}$ tetrahedral clusters and have magnetic frustration effects~\cite{Diepbook,PhysRevLett.93.217206,PhysRevB.71.224430,PhysRevB.79.014406}. They crystallize in the noncentrosymmetric tetragonal SG \#81, and four Cu$^{2+}$ ions lying closest
to each other form a distorted tetrahedron.
The spins of the Cu$^{2+}$ ions are $S = \frac{1}{2}$, and show incommensurate long-range ordering at low temperatures ($T_C=18.2$ and 11.4 K for $X=$ Cl  and Br, respectively)~\cite{PhysRevLett.93.217206}. In this work, we focus on the paramagnetic (PM) phase above $T_C$. Taking \CuBr~as an example in the main text, the PM band structure of \CuBr~is presented in Fig.~\ref{fig:152959_arc}b. Based on the irreps on the maximal high-symmetry $k$-points, its occupied bands satisfy the CR and have a trivial SI. Accordingly, there is always a direct band gap along the high-symmetry lines, as shown in Fig.~\ref{fig:152959_arc}b and its inset. It was previously classified as a trivial insulating phase.
However, from the plotted WCC along $\wt\Gamma$--$\wt M$ in Fig.~\ref{fig:152959_arc}a, the $\chi$ is read to be $-2$, which implies that the PM \CuBr~is a WSM.  Consequently, 24 Weyl points are found in the PM phase. Their positions, topological charges and energies are tabulated in Table~\ref{tab:charge}. On (100) surface, the constant surface energy contours are obtained for $E-E_{\rm F}=0.0989$ eV (Fig.~\ref{fig:152959_arc}c) and $E-E_{\rm F}=0.1140$ eV (Fig.~\ref{fig:152959_arc}d). The surface Fermi arcs are visible and terminated at the projections of the Weyl points. Therefore, the compounds Cu$_2$Te$_2$O$_5X_2$ serve as good platforms for studying the interplay between Weyl points and magnetic frustration effects in the interacting spin-$\frac{1}{2}$ tetrahedral clusters.

\begin{figure*}[!t]
\includegraphics[width=16.8cm]{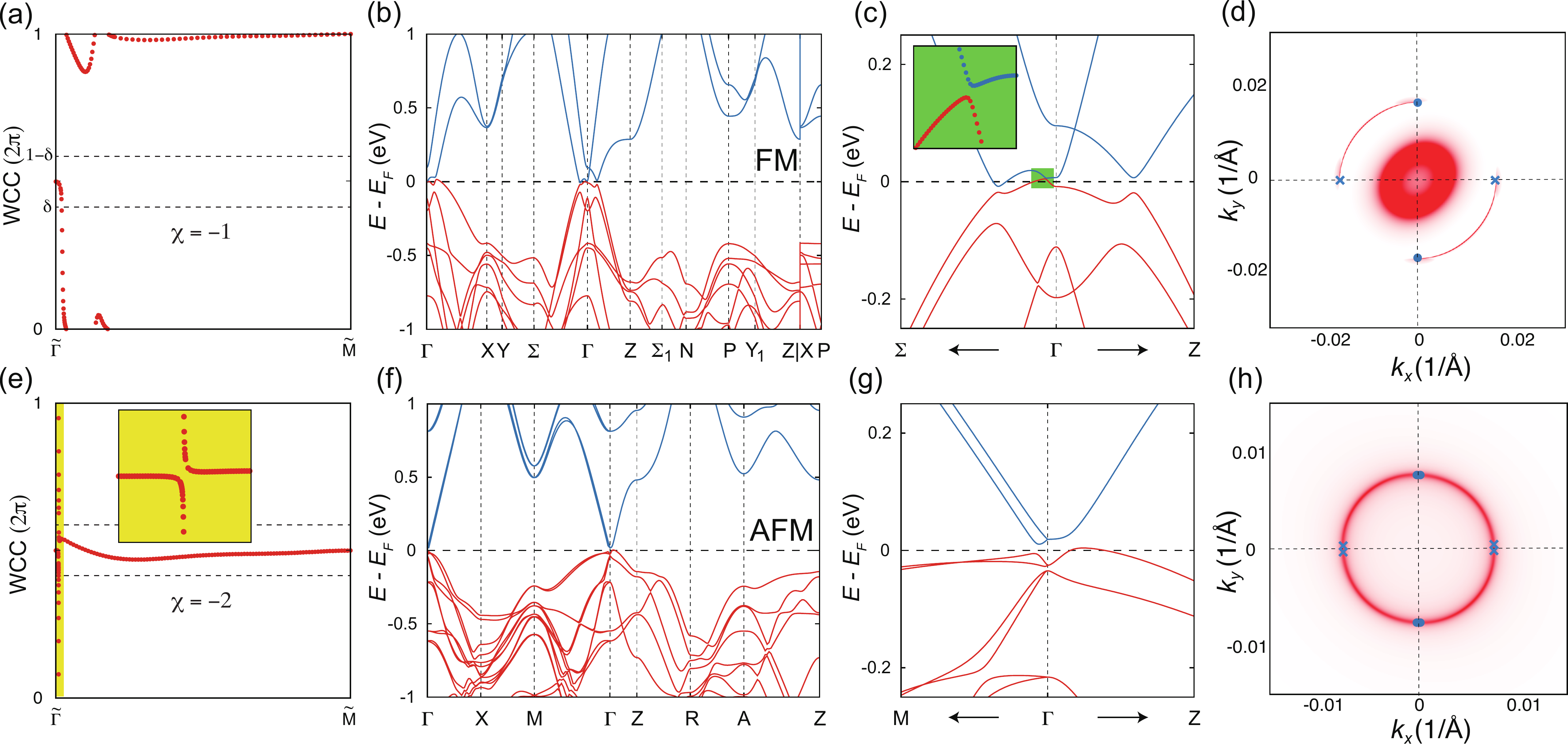}
\caption{(Color online)
(a) and (e) The Wilson bands of the $xy$-plane path of FM and AFM \mag, respectively. (b) and (f) The band structures of FM and AFM \mag, respectively.
(c) and (g) The zoom-in band structures around $\Gamma$ point in (b) and (f), respectively.
(d) and (h) The (001) surface states of FM \mag~at $E-E_{\rm F}=0.0057$ eV and AFM  \mag~at $E-E_{\rm F}=0.01$ eV, respectively.
} \label{fig:mag}
\end{figure*}

\subsubsection{The class of nonmagnetic WSMs with nontrivial SIs}

On the other hand, the obtained nonzero invariant $\chi$ also suggests that some materials, previously classified as topological insulators based on nontrivial SIs, turn out to be WSMs. For this class of WSMs, we mainly introduce the A$_3$B compounds with the Ni$_3$P structure, especially Zr$_3$Sb and Ta$_3$Ge.  As we know, A$_3$B are a very large family of well-known intermetallic compounds with Bardeen-Cooper-Schrieffer superconductivity, attracting a lot of attention due to the high superconducting transition temperature over decades~\cite{stewart2015superconductivity,muller1980a15}. They usually crystallize in three different structures: cubic A15 structure~\cite{stewart2015superconductivity}, tetragonal Ti$_3$P-type and Ni$_3$P-type structures~\cite{waterstrat1975refinement,willis1978superconductivity}.
Here, we focus on the Zr$_3$Sb and Ta$_3$Ge of the Ni$_3$P-type structure with SG \#82, which become superconducting below 1 K~\cite{willis1978superconductivity}.
In this work, we find that they are actually WSMs in their normal state, even though they satisfy the CR in SG \#82.

In the databases discovered by the symmetry-based strategy~\cite{tang2019comprehensive,zhang2019catalogue,vergniory2019complete}, the S$_4$ $z_2$ in Zr$_3$Sb and Ta$_3$Ge is computed to be 1, leading that Zr$_3$Sb and Ta$_3$Ge are classified as the topological insulating phase.
Although there is a continuous direct band gap in the band structure of Fig.~\ref{fig:195057_arc}b, the $k_z$-directed Wilson-loop calculations reveal the nontrivial topological invariant $\chi=-6$ for Zr$_3$Sb and $\chi=-10$ for Ta$_3$Ge, as shown in Fig.~\ref{fig:195057_arc}a and the Supplementary materials, respectively, guaranteeing the appearance of Weyl points in these materials. The Fermi arc states of Zr$_3$Sb on (100) surface are shown in constant energy contours in Fig.~\ref{fig:195057_arc}c and d for $E-E_{\rm F}=-0.0673$ eV and $E-E_{\rm F}=0.0906$ eV, respectively.
Given the coexistence of superconductivity and Weyl points in these noncentrosymmetric systems, they are good platforms for studying the potential 3D topological superconductivity when the states originating from the Weyl points open superconducting gaps~\cite{qi2010topological,hosur2014time,qian2019topological}.

\subsubsection{The class of magnetic WSMs}
In addition, the magnetic WSMs can be indicated by $\chi\neq 0$, since the topological invariant $\chi$ is applicable to the systems without TRS.
By employing the Wilson-loop technique, the possible WSM phases in the I$_2$-${\text{II}}$-${\text{IV}}$-VI$_4$ compounds are checked for both ferromagnetic (FM) and antiferromagnetic (AFM) spin configurations, where I= Cu, Ag; II= Mn, Fe, Co, Ni; IV= Si, Ge, Sn; and VI= S, Se, Te.
Magnetic semiconducting compounds by replacing the II cations with Mn$^{2+}$, Fe$^{2+}$, Co$^{2+}$ and/or Ni$^{2+}$ ions are of interest because of the large magneto-optical effects and the manner, in which the magnetic behavior associated with the concerned magnetic ions can modify and complement the semiconductor properties.  In these compounds, a variety of magnetic states can occur, such as FM state, AFM state, canted FM state, and magnetic frustration due to magnetic competition between magnetic neighbors, etc. Here, based on first-principles calculations with spin polarization, we find that as an illustration example, the Cu$_2$MnSnSe$_4$ compound of the tetragonal stannite structure hosts Weyl points in both FM and AFM states.

The band structures of FM and AFM \mag~are shown in Fig. \ref{fig:mag}b and f, respectively. Although the conduction bands are very close to the valence bands, the zoom-in band structures clearly show that there is a direct band gap along the high-symmetry lines for FM and AFM \mag, as shown in Fig. \ref{fig:mag}c and g, respectively.
In order to figure out their bulk topology, the WCC of $k_z$-directed Wilson loops of them are calculated and presented in Fig. \ref{fig:mag}a and e, respectively. The topological invariant is computed to be $\chi=-1$ for the FM state and $\chi=-2$ for the AFM state, indicating the existences of Weyl points in both states. By carefully checking the energy gaps and the associated chiral charges, we find 12 and 8 Weyl points in FM and AFM \mag, respectively. The positions, charges, energies and multiplicities of these Weyl points are shown in Table \ref{tab:charge}. The existences of Fermi arc surface states (Fig. \ref{fig:mag}d and h) also confirm that both FM and AFM \mag~are WSMs.
So the nontrivial topological invariant ($\chi=-1$) is consistent with the indicator $s_4=1$, which can guarantee Weyl points in magnetic \mag.
In experiments, AFM, FM and spin-glass behaviors of Cu$_2$MnSnSe$_4$ have been reported ~\cite{guen1980electrical,chen1993magnetic,quintero2010magnetic}. So it is expected to observe Weyl points in magnetic \mag~in future experiments.

\section{Conclusion}
We propose a new topological invariant $\chi$ for the S$_4$-invariant systems, which can be computed through the 1D Wilson-loop technique.
By computing the new invariant $\chi$ in our first-principles calculations, we have performed high-throughput screening for WSMs in materials of 20 S$_4$-invariant SGs.
The new method of the invariant $\chi$ is very efficient and can be widely used to predict WSMs with S$_4$ symmetry in both magnetic and nonmagnetic materials. A lot of WSMs are predicted theoretically in this work and the Fermi-arc surface states for some representatives are presented as well.
Many predicted WSMs, in which various interesting properties (e.g., magnetic frustration effects, superconductivity or spin-glass order, etc.) are found, provide realistic platforms for studying the interplay between Weyl fermions and other exotic states in future experiments.

\ \\

\noindent \textbf{Conflict of interest}
The authors declare that they have no conflict of interest.
\ \\

\noindent \textbf{Acknowledgments}
This work was supported by the National Natural Science Foundation of China (11974395, 11504117, 11674369, 11925408), the
Strategic Priority Research Program of Chinese Academy of Sciences (XDB33000000) and the Center for Materials Genome. 
H.W. acknowledges support from the National Key Research and Development Program of China 
(2016YFA0300600, 2016YFA0302400, and 2018YFA0305700), and the K. C. Wong Education Foundation (GJTD-2018-01).

\ \\
\noindent \textbf{Author contributions }
Zhijun Wang and Simin Nie proposed and supervised the project. Jiacheng Gao and Yuting Qian carried out DFT calculations. Jiacheng Gao, Simin Nie and Zhijun Wang did the theoretical analysis. All authors contributed to writing the manuscript.


\end{document}